\documentclass[]{emulateapj}

\begin{document}

\title{Modeling Spitzer observations of VV Ser. II. an extended quantum heated nebula and a disk shadow}

\author{Klaus M. Pontoppidan\altaffilmark{1,2,5}}

\email{pontoppi@gps.caltech.edu}
\altaffiltext{1}{California Institute of Technology, Division of Geological and Planetary Sciences, MC
150-21, Pasadena, CA 91125}
\altaffiltext{2}{Leiden Observatory, P.O.Box 9513, NL-2300 RA Leiden, The Netherlands}

\author{Cornelis P. Dullemond\altaffilmark{3}}
\altaffiltext{3}{Max-Planck-Institut f{\"u}r Astronomie, K{\"o}nigstuhl 17, 
Heidelberg, 69117, Germany}

\author{Geoffrey A. Blake\altaffilmark{1}}

\author{Neal J. Evans II\altaffilmark{4}}
\altaffiltext{4}{Department of Astronomy, University of Texas at Austin, 1 University Station, C1400, 
Austin, TX 78712-0259}

\author{Vincent C. Geers\altaffilmark{2}}

\author{Paul M. Harvey\altaffilmark{4}}

\author{William Spiesman\altaffilmark{4}}
\altaffiltext{5}{Hubble Fellow}

   \begin{abstract}

We present mid-infrared Spitzer IRAC and MIPS images of the UX Orionis
star VV Ser and the surrounding cloud. The 5.6--70\,$\mu$m images show bright, localized and nebulous emission
extended over $4\arcmin$ centered on VV Ser. We interpret the nebulosity
as being due to transiently heated grains excited by UV photons emitted by VV Ser.
A companion paper describes how the physical structure of the 
VV Ser disk has been constrained using a wide range of observational data modeled by an axisymmetric 
Monte Carlo radiative transfer code. The model of the system in particular constrains 
the strength of UV radiation field throughout the envelope, and the nebulosity thus presents an opportunity for
directly constraining the excitation mechanism and opacity law of PAHs in molecular clouds. In this paper we employ the
model to study the nebulosity surrounding VV Ser using quantum-heated PAH molecules
and Very Small Grains (VSGs) consisting of amorphous carbon 
in the thermal cooling approximation.
Imprinted on the nebulosity is a wedge-shaped dark band, centered on the star. We
interpret this dark wedge as the shadow cast by the inner regions of
a near-edge-on disk in UV light, allowing the PAHs to be excited only
outside of this shadow. The presence of a disk shadow strongly constrains the inclination as well as the
position angle of the disk. The nebulosity at 5.6--8.0\,$\mu$m and 
the $2175\,\rm\AA$ absorption feature seen in an archival spectrum from the International Ultraviolet Explorer 
can be fit using only PAHs, consistent with the main carrier of
the $2175\,\rm\AA$ feature being due to the graphite-like structure of the PAHs. The PAH component
is found to be relatively smoothly distributed in the cloud, while the population of VSGs emitting at 20--70\,$\mu$m
is strongly concentrated $\sim 50\arcsec$ to the south east of VV Ser. Although depending on the adopted PAH opacity, the
abundance of PAHs in the surrounding cloud is constrained to $5\pm2\%$ of the total dust mass, given the opacity. 
The extent of the nebulosity constrains the density of the gas surrounding the
VV Ser disk to $500\pm200\,\rm cm^{-3}$ for a gas-to-dust ratio of 100. This low density suggests that the quantum heated material
is not part of the original envelope of VV Ser and that it is rather a quiescent part of the Serpens molecular cloud that
the star has passed into after being formed. 
Although relatively rare, quantum 
heated nebulosities surrounding single, well defined stars are therefore well-suited for gaining unique insights into
the physics of very small particles in molecular clouds. 

   \end{abstract}

\keywords{accretion, accretion disks -- circumstellar matter 
-- stars: formation, pre-main-sequence -- infrared: stars }

\section{Introduction}

It is well known that if a sufficiently small grain is hit by a high energy (ultraviolet) photon, 
the internal energy content far exceeds that of a 
grain in thermal equilibrium \citep{Duley73, Sellgren84}. Once the small grain has been heated, it will rapidly 
cool by emitting strongly in the mid-infrared wavelength regime. 
Yet, because the energy is distributed in many vibrational modes (typically $3(N_{\rm atom}-2)$), the
subsequent cooling of the grain takes place in an almost thermal (classical) manner.
Such quantum-heated grains have been used to explain a range of observations in the past. Specifically, 
emission from polycyclic aromatic hydrocarbons (PAHs) require quantum-heating \citep{Leger84,Siebenmorgen92}, but also
the presence of very small carbon/silicate grains (VSGs) of sizes $5-50\,\mathrm{\AA}$ 
has been inferred by infrared observations \citep{Boulanger88,Henning98}.
In many extragalactic environments, observations with the {\it Infrared Space Observatory} (ISO) have
shown the importance of emission from quantum-heated grains. 
VSGs have been suggested to play a crucial role for the emission spectra of especially starburst galaxies 
\citep{Contursi01,Schreiber03,Rouan04}. Typical for all environments with high UV fields is that PAHs dominate
the spectrum from 5--15\,$\mu$m, while VSGs dominate at wavelengths $\gtrsim 15\,\mu$m.

This paper is the second part of a study of Spitzer observations of the
Herbig Ae star VV Ser. We focus on VV Ser for several reasons. 
Mid-infrared images obtained with the {\it Spitzer Space Telescope} as part of the Legacy Program 
``From Molecular Cores to Proto-planetary Disks'' \citep{Evans03} show a bright bipolar nebulosity surrounding
the central star. 
This was a considerable surprise since the nebulosity has no or little
optical and near-infrared counterparts and therefore the density must be low such as not to create any
detectable reflection nebulosity. The simplest way of explaining this, is if the nebulosity observed by
{\it Spitzer} is caused by very small ($\lesssim 0.01\,\mu$m) transiently heated dust grains. 

Another important reason for selecting VV Ser for further study is the presence of a dark band bisecting
the entire mid-infrared nebulosity.
The dark band has a ``wedge-like'' shape suggesting the presence of
a nearly edge-on circumstellar disk. Yet, the extent of the band of $>4\arcmin$ or 60,000\,AU 
is far too large for a circumstellar disk.  
Exactly this type of morphology has been seen in reflection nebulosities surrounding a number
of young stars \citep{Hodapp04, Pontoppidan_shadow}. These studies suggest that if an edge-on or nearly 
edge-on disk is surrounded by large-scale envelope
material (or diffuse material), then the shadow cast by the disk into/onto
the surrounding material in scattered light can be seen as a dark wedge-like
band originating from the star and extending out to tens of thousands of AU,
i.e. many times the size of the disk. 
There are good reasons to believe that any extended PAH emission excited by a single source should 
exhibit a similar behaviour due to the shadowing of ultraviolet photons by a circumstellar disk. 
In fact, the effect may be more pronounced in the case of quantum-heated grains 
since the exciting UV radiation from the
central star will be subject to a stronger shadowing effect than near-infrared photons. 
The opening angle of a disk shadow and the sharpness of the shadow may constrain the structure and
composition of the disk as well as the opacity spectrum of the quantum-heated grains. 
Disk shadows are therefore a powerful tool for studying the physical structure of
spatially unresolved circumstellar disks. Clearly, the best way of analyzing these data is the
simultaneous fitting of all available data with radiative transfer models of
disks surrounded by an envelope.

There is considerable additional evidence that the VV Ser system is highly inclined ($\gtrsim 70\degr$).
This includes broad, double-peaked CO rovibrational emission lines from the fundamental band around $4.7\mu$m 
\citep{Blake04}. 
Near-infrared interferometric measurements, although somewhat ambiguous, seem to point to a high inclination 
\citep{Eisner03,Eisner04}. Most importantly, perhaps, VV Ser is a well-known UX Orionis variable star, i.e.
the star experiences frequent, but irregular, brightness dips, presumably due to orbiting dust clumps in the disk passing in front
of the star.
The first part of this study \citep[][henceforth Paper I]{paperI}
describes the observations of VV Ser in terms of its status as a UX Orionis type
variable and argues extensively for a high inclination of the VV Ser disk. 
In Paper I we construct an axisymmetric Monte Carlo radiative transfer model
of the VV Ser disk and use the model to show that the Spectral Energy Distribution (SED), the near-infrared interferometric
visibilities of \cite{Eisner04} as well
as the UX Orionis variability can be explained in terms of a self-shadowed nearly edge-on
disk. In the present study we use the same model with the 
addition of quantum heated grains to study the large extended nebulosity surrounding
VV Ser. Our goal is to construct a single radiative transfer model that explains all the observed
characteristics of VV Ser in terms of an inclined disk embedded in a relatively low density cloud containing
PAHs and VSGs. Once the structure of the system as well as the radiation field is known, the physical properties of 
the quantum heated grains surrounding VV Ser can be constrained, in particular their abundance, and
to some degree, their opacities. 

This article is organized as follows: In Sect. \ref{Obs} we describe the {\it Spitzer}
imaging observations of VV Ser and in Sect. \ref{Cloud} the cloud environment in
which VV Ser is embedded. Sect. \ref{Model} describes the radiative transfer model
and our treatment of quantum-heated grains. Sect. \ref{Features} discusses how the
{\it Spitzer} images and other imaging data compare to the model and Sect. \ref{Discussion} provides a discussion 
in particular of how quantum-heated nebulosities around isolated low-mass stars can be used to
constrain the small grain population associated with star formation.

\section{Observations}
\label{Obs}

Most of the observations of VV Ser used in the modeling are described in Paper I. 
The mid-infrared spectroscopy and imaging
were obtained with the {\it Spitzer Space Telescope} \citep{Spitzer} as part of the Legacy program, {\it ``From
Molecular Cores to Protoplanetary Disks''} (c2d) \citep{Evans03}, using all the available
instruments [IRS, \cite{IRS}, IRAC, \cite{IRAC} and MIPS, \cite{MIPS}]. In addition to the 
Spitzer-IRS mid-infrared spectroscopy of VV Ser presented in Paper I
we use IRAC and MIPS imaging at 3.6, 4.5, 5.6, 8.0, 24 
and 70\,$\mu$m of the area surrounding VV Ser. The imaging was reduced using the c2d 
mosaicking pipeline (see \cite{Harvey06}; Spiesman et al. in prep. for IRAC and MIPS,
respectively).
In support of the Spitzer observations, we have also obtained a continuum map
at 850\,$\mu$m with SCUBA on the James Clerk Maxwell Telescope (JCMT) (see also Paper I). 
From the European Southern Observatory archive, we have extracted archival
images in the $J$-, $H$- and $K$-bands obtained with SOFI at the New Technology
Telescope (NTT)\footnote{In part based on observations obtained at the European Southern Observatory at La Silla
under program 67.C-0042(A).}. The near-infrared images were reduced using standard methods of
dark subtraction, flat field correction and registering of individual frames using
cross correlation. An ultraviolet spectrum obtained with the International 
Ultraviolet Explorer (IUE) was obtained from the IUE archive and is described in Paper I. 

\begin{figure*}
  \plottwo{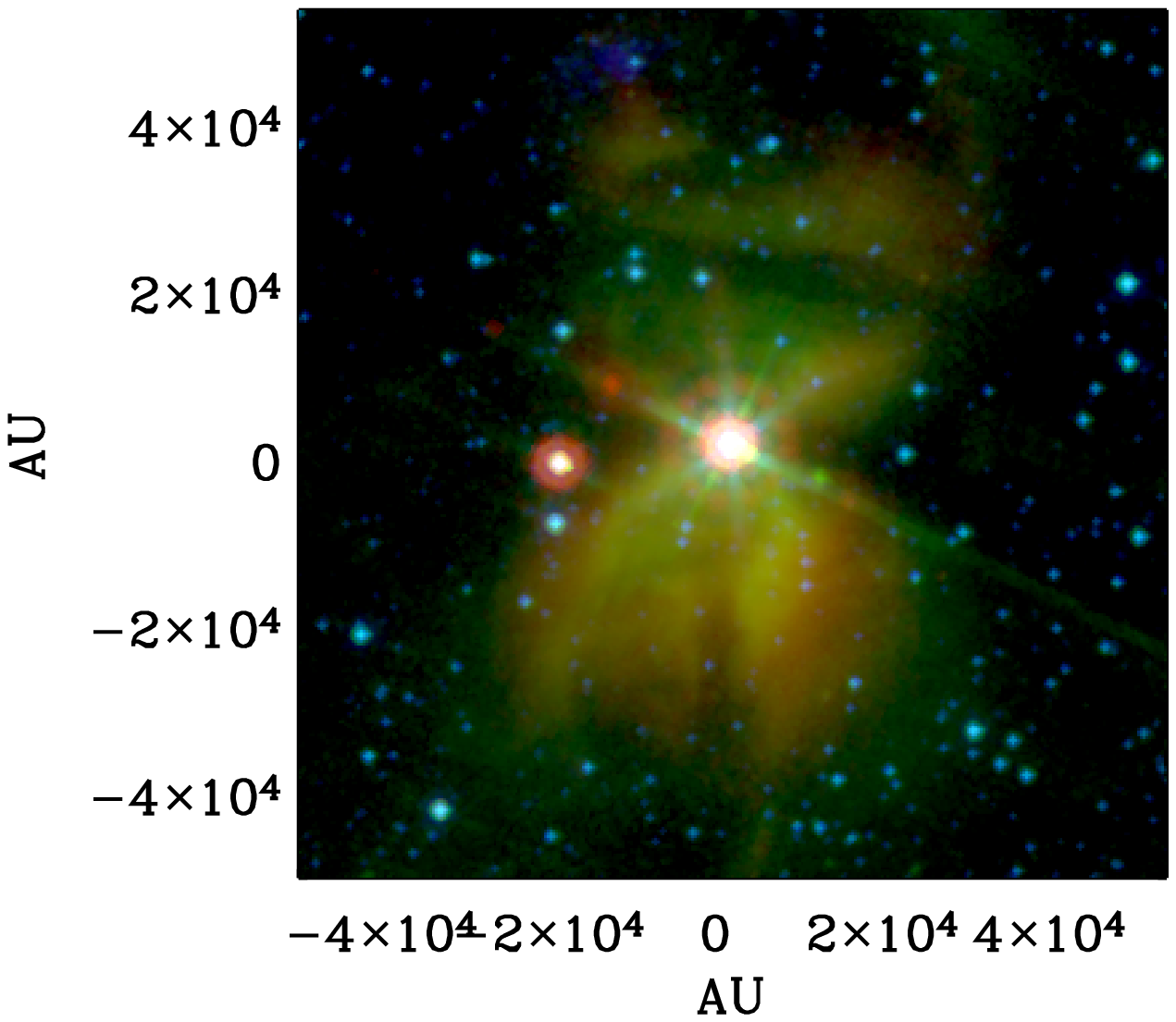}{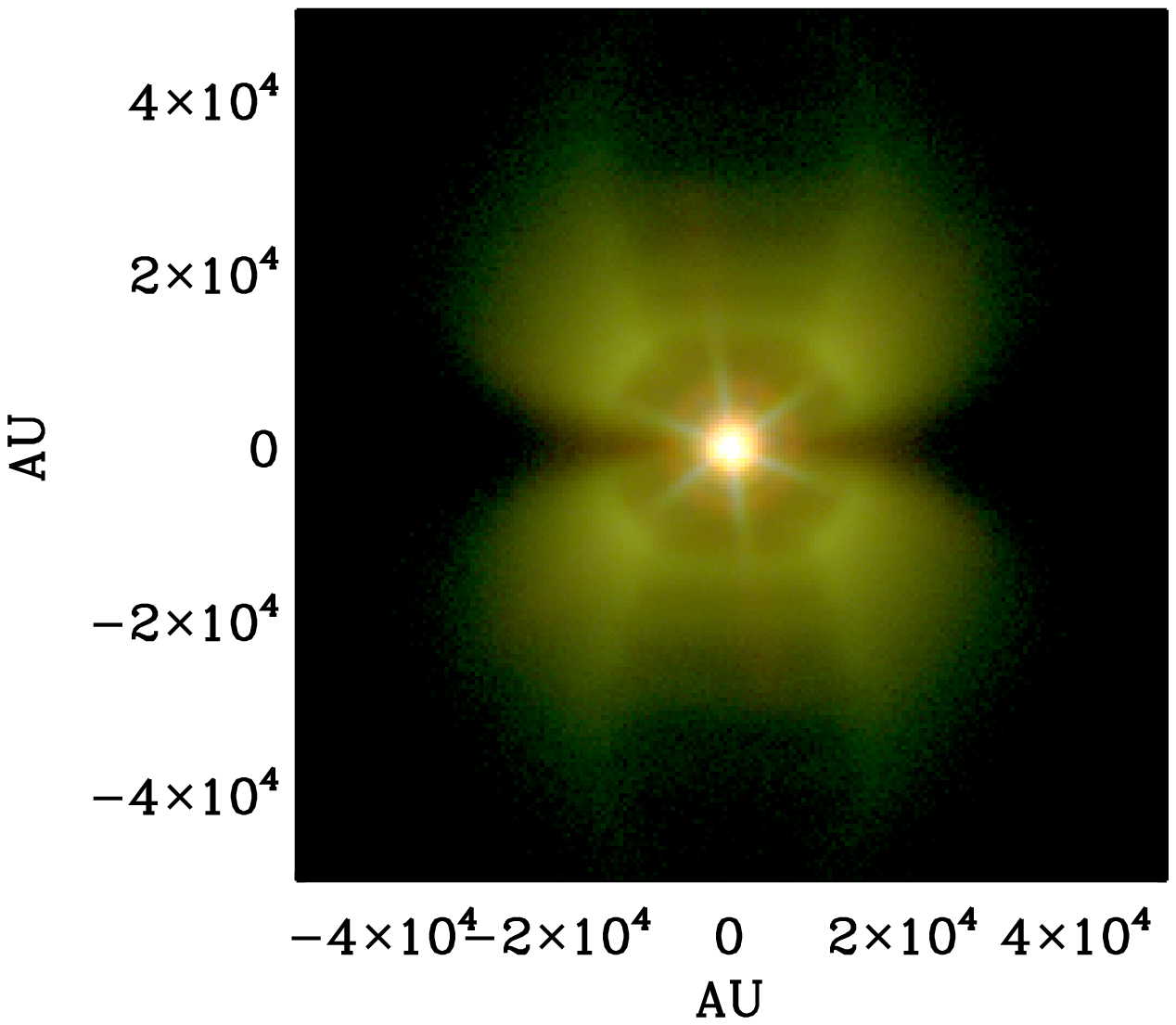}
  \caption{4.5 (blue), 8.0 (green), 24.0\,$\mu$ (red) color composites of the
VV Ser nebulosity. {\it Left panel:} Spitzer-IRAC bands 2 and 4 together
with Spitzer-MIPS band 1. The field is centered on VV Ser and measures 
$400\arcsec$ on the side. The bright red source to the left of VV Ser is a 
presumably unrelated young star, a background evolved star or possibly a very bright background infrared
galaxy. In this color scheme, background stars are blue, 
while PAH emission is green. {\it Right panel:} Model color composite. 
The model has been convolved with theoretical PSF profiles calculated using the STINYTIM
package ver. 1.3 for pre-launch parameters (Krist 2002, SSC web site). 
Additionally, Gaussian noise has been added to the model
with a standard deviation corresponding to that of the actual Spitzer images.}
  \label{Images}
\end{figure*}

\section{The molecular surroundings of VV Ser}
\label{Cloud}

VV Ser was identified as a particularly interesting object when mid-infrared 
IRAC and MIPS images revealed that the star is surrounded by a bright, extremely large 
nebulosity not seen in near-infrared images. An IRAC/MIPS color composite image is shown 
in Fig. \ref{Images}. The nebulosity extends over an
area at least 6 arcminutes (or 94,000\,AU at a distance of 260\,pc) 
across and is brightest in the 8.0, 24 and 70\,$\mu$m bands (IRAC4, MIPS1 and MIPS2), 
but is also weakly detected in the 5.6\,$\mu$m band (IRAC3).
At 3.6 and 4.5\,$\mu$m (IRAC1 and IRAC2), no extended emission is detected. 
We interpret the nebulosity as being due to a quantum-heated component of either an
envelope or simply the surrounding molecular cloud. The relatively low extinction ($A_V\sim 3\,$mag) 
toward VV Ser strongly suggests that little original envelope material is left. This quantum-heated 
component likely consists partly of Polycyclic Aromatic Hydrocarbons (PAHs) 
and partly of Very Small Grains (VSGs). The presence of the VSGs is inferred from the bright
24 and 70\,$\mu$m nebulosity (see Sect. \ref{nebulosity}). 
The presence of such compact, mid-infrared nebulosities is clearly a rare, but not unique,
phenomenon, judging from the larger {\it c2d} Spitzer maps. 

The optical colors of the star 
correspond to a steady extinction of $A_V \sim 3$, but with frequent, 
non-periodical dips
lasting of order 10 days of 0.5--3.0 additional magnitudes of extinction. 
Using the 2MASS point source catalog, an average extinction through the
Serpens cloud in a $5\arcmin\times 5\arcmin$ field around VV Ser is 
determined by fitting a reddened main sequence to the colors of background
stars in the field. Assuming a near-infrared extinction law of 
$A_{\lambda}\propto \lambda^{-1.9}$ \citep{Kaas99} an average extinction
of $A_J=1.5\pm 0.2$ is found. For an $R_V=3.1$ optical extinction law, this corresponds to
an $A_V\sim 6$. The lower resolution extinction map of \cite{Cambresy99} finds an extinction of
$A_V=7-8$\,mag in the region. For a constant density molecular cloud, this places VV Ser roughly in the middle
(along the line of sight) of a filament aligned in the north-south direction. The northern end of
the filament culminates in the famous Serpens Cloud Core [incidentally, the Serpens Core contains CK 3, 
another disk shadow candidate, \cite{Pontoppidan_shadow}]. 
The width of the filament is roughly $22\arcmin$ \citep{Cambresy99}. 
Using the conversion between column density and extinction in the $J$-band from \cite{Vuong03},
$N(\mathrm{H}_2)=2.8\times 10^{21}\,\mathrm{cm}^{-2}\times A_J$, the molecular cloud density
around VV Ser is estimated to $\sim 700\,$cm$^{-3}$.

\section{Model}
\label{Model}
To model the observed SED and Spitzer imaging of VV Ser, we use the
axisymmetric Monte Carlo radiative transfer code RADMC \citep{Dullemond04}
in combination with the raytracing capabilities of the more general code RADICAL 
\citep{Dullemond00}. The density structure
is axisymmetric, but the photons are followed in all three dimensions.  
The RADMC code is used to derive the temperature structure and scattering source function of a given dust 
distribution. Once these parameters are known, images can be calculated using RADICAL. 
This setup has been used to model similar protostellar disks
\citep{Pontoppidan_crbr, Pontoppidan_shadow}. In Paper I it is described how the
physical disk structure is constrained to fit the observed SED, images and UX Orionis behaviour.
In this paper, we add a simple treatment of quantum-heated
grains to the model in order to describe the mid-infrared nebulosity 
surrounding VV Ser. 
We note that while the treatment of quantum heated grains included in the model
is intended for the VV Ser nebulosity, it can easily be adapted to treat a wide range
of  axisymmetric radiative transfer problems. One obvious possibility is the treatment of
emission from PAH molecules in the disk itself, although that would require the
inclusion of multi-photon excitations. The code only treats isotropic scattering, which is not a good
approximation for the UV photons considered in the models. However, in terms
the quantum heated nebulosity we argue that the scattering phase function only plays a minor role. 
Since the optical depth to scattering of UV photons is only $\sim 0.5$ through the envelope, the majority
of photons heating small grains in the extended nebulosity have experienced at most a single scattering event
from the disk surface.
This introduces an error of up to a factor of two in surface brightness - similar to the uncertainty on the 
opacities used.

\subsection{Quantum heated grains and multi-dimensional radiative transfer}
\label{QH}
Quantum heated grains are treated in the model using the 
``continuous cooling'' approximation \citep[e.g.][]{Guhathakurta89}. 
This popular approximation is valid for sufficiently large grains/molecules.
In principle, it should only be used for grains with more than 100 atoms, but in the case of PAHs it
was found by \cite{Draine01} to provide accurate results for molecules as small as C$_{30}$H$_x$ by comparing to the full statistical treatment.
The continuous thermal cooling approximation is a method for calculating PAH emission spectra
without having to treat the radiative transfer in a full quantum mechanical framework.
When including quantum heating in a multi-dimensional radiative transfer code, this approximation
results in an significant advantage in terms of CPU time relative to a code calculating a full transition matrix treating
all upwards and downwards transitions between vibrational modes. 
We emphasize that the code presented here is also very portable to any kind of axisymmetric structure.

In the adopted approximation, 
grains are heated in a quantized fashion, but {\it cooled classically}. The necessary assumption is that a
sufficient number of vibrational modes are excited in the grain to create a continuous cooling behaviour, at
least in the temperature range where most of the power is radiated. In terms of the radiative transfer, a
necessary assumption is that the absorbed and emitted radiation fields are decoupled. In the case of PAHs, this amounts
to assuming that infrared photons do not excite the PAH molecules to significant temperatures. 
In the current version of the code, only single photon excitations are treated.
Multi-photon excitations are important for very strong radiation fields and do not apply to
the nebulosity surrounding VV Ser at distances of 5,000--50,000\,AU from the central star, where
no additional photons will strike a given PAH molecule in the time it takes to cool to the background temperature. For 
inclusion of disk PAHs, multi-photon excitations will be essential and therefore we do not attempt to model
the disk PAH features seen in the {\it Spitzer} spectrum. 

\subsection{The thermal cooling approximation}
The treatment of quantum heated grains can in the thermal cooling approximation be regarded as a post-processing add-on
to the radiative transfer Monte Carlo code. The Monte Carlo code uses the \cite{Bjorkman01} method
with continuous absorption according to the \cite{Lucy99} algorithm. This method
follows photon packages from the surface of the central star as they travel through
a gridded density structure. Photon packages can be
scattered, changing their directions, or absorbed, altering both their
wavelength and direction. The code supplies the UV radiation field at each grid
point, and the calculation of the non-thermal emission is a matter of calculating a response function that can convert
the fraction of UV photons absorbed by non-thermal grains into infrared photons.

In essence, to avoid the computationally intensive
task of re-calculating the temperature distribution function of the quantum heated grains every time a photon interacts with a grid cell, we
split the problem in the following way: The mean intensity of quantum heating photons is obtained
by adding the energy absorbed from each photon packet passing through a given cell. The mean intensity for each grid
point is then saved and used at the end of the Monte Carlo run to calculate the source function of quantum heated grains, as
described below. To facilitate this approach, the photon packages are divided into primary and secondary
components. The primary photon packages are those that originate at the stellar surface, and are
only scattered. If a primatry photon package is absorbed by a quantum-heated grain, it is lost to
the Monte Carlo code. If it is absorbed and re-emitted by a thermal grain, it is converted to a secondary
photon package. All grains, including quantum-heated grains, are treated as thermal
by a secondary photon package. After completing the Monte Carlo run, the primary photon packages 
absorbed by the quantum-heated grains are used by an external code to calculate a quantum-heated emissivity. 
When this is done, a second Monte Carlo run can be executed in which
photons can be launched from the quantum heated grains according to the emissivity 
calculated by the external code, in addition to photons launched from the stellar surface. Photon packages
launched by quantum heated grains begin as secondary grains, and will only heat thermal
grains. Typically, the effects of this second Monte Carlo run on the SEDs and images are minor. 

For PAHs, this split between primary and secondary photon packages is likely accurate since the 
PAH opacity drops by at least 3 orders of magnitude above 1\,$\mu$m, according to the
opacities of \cite{Li01}. So even though there are 10--100 times more photons at 
near-infrared wavelengths than at UV wavelengths, only 1-10\% of the heating events will potentially be due to infrared photons.
However, recent laboratory results have shown that at least some classes of ionized PAHs may have significant cross sections 
in the near-infrared \citep{Mattioda05}. Clearly, future work will need to explore the effects of this in further detail.
In terms of the model, an under-estimate of the near-infrared cross section may result in an over-estimate of the
abundance of quantum heated grains. Note, however, that single infrared photons will not be able to excite
PAHs to the temperatures responsible for the short-wavelength PAH features below ~10\,$\mu$m. 
 For carbonaceous or silicate VSGs, this assumption may be less valid because these types of grains have a less steep opacity
curve. 

In the following, $\bar{\nu}$ is a frequency of the UV radiation field (primary) photon as opposed to $\nu$, 
which indicates a frequency of a re-emitted secondary photon.

The cooling rate of the quantum heated grains, assumed to be radiating thermally, is:
\begin{equation}
\Gamma_{\rm cool} = \int_{0}^{\infty}4\pi B_{\nu}(T) \kappa_{\nu}d\nu,
\end{equation}
where $\kappa_{\nu}$ is the opacity of the quantum heated grains, and $B_{\nu}(T)$ is the Planck function.
The time to cool to a temperature, $T$, can then be calculated as:
\begin{equation}
\tau_{\rm cool}(T) = \int_{T}^{T_{\rm peak}}\frac{C(\bar{T})}{\Gamma_{\rm cool}(\bar{T})} d\bar{T},
\label{CoolingCurve}
\end{equation}
where $C(T)\equiv dU/dT$ is the heat capacity of the dust material. $U(T)$ is the enthalpy curve related to the
bulk material of which the quantum heated grains consist. The grain contains $N_{\rm atom}$ atoms, which
in the case of a PAH molecule refers to the number of carbon atoms. The peak temperature, $T_{\rm peak}$, reached
when the grain is hit by a photon of energy $\bar{\nu}$ is simply found via the enthalpy curve:
\begin{equation}
U(T_{\rm peak})=h\bar{\nu}/N_{\rm atom}
\end{equation} 
The next step is to calculate the time-integrated emission per mass of the quantum-heated grains. 
This produces the output energy spectrum emitted if one photon of energy $h\bar{\nu}$ is absorbed.
\begin{equation}
E_{\nu}(\bar{\nu}) = \int_{T_{\rm min}}^{T_{\rm peak}}\kappa_{\nu} B_{\nu}(T)\times \left( \frac{d\tau_{\rm cool}}{dT}\right) dT,
\end{equation}
The emissivity of the quantum heated grains is then determined by the UV radiation field alone, $\chi_{\bar{\nu}}$ (e.g. in units
of erg\,s$^{-1}$\,cm$^{-2}$\,Hz$^{-1}$\,sterad$^{-1}$), and the mass density of transiently heated particles, $\rho_{\rm QH}$.
Integrating over the photons absorbed by the quantum heated grains defines a response matrix, yielding an 
emissivity $j_{\nu}$ as a function of an input radiation field spectrum:
\begin{equation}
j_{\nu}(\chi_{\bar{\nu}}) = \int_{0}^{\infty} \beta_{\bar{\nu}}\kappa_{\bar{\nu}}\rho_{\rm QH}\frac{\chi_{\bar{\nu}}}{h\bar{\nu}} E_{\nu}(\bar{\nu})d\bar{\nu}
\end{equation}

We have included the effect of destroying the grains if $T_{\rm peak}$ is higher than the sublimation temperature. 
In essence this works as an energy sink since the absorbed energy is assumed to be used to heat the gas and
subsequently be emitted as gas-phase lines in an optically thin part of the spectrum.
The grain destruction is described by the step function, $\beta_{\nu}$:

\begin{equation}
\beta_{\bar{\nu}} = \left\{ \begin{array}{ll} 
1 & \mathrm{for}\ T_{\rm peak} \leq 1500\,\rm K \\
0 & \mathrm{otherwise}
\end{array} \right. 
\end{equation}
In fact, for the single photon excitation case, no grains are destroyed, since a 0.1\,$\mu$m photon will excite a C$_{50}$H$_{20}$
PAH molecule to just under 1500\,K and a 1000 atom silicate grain to 200\,K. However, for the multi-photon excitation
case, grain destruction may play an important role for the radiation fields present within a few AU of a Herbig AeBe star.

\subsection{PAH and VSG opacities and enthalpy curves}

The physics of a given population of quantum heated grains is contained in the opacity, in the enthalpy curves and the grain size.
The presence of strong 24 and 70\,$\mu$m nebulosity in addition to the 6--8\,$\mu$m emission suggests
that both a PAH component (providing the short wavelength opacity) as well as a carbon and/or silicate VSG component (to provide
the far-infrared opacity) are required. 
The PAH opacity, $\kappa_{\rm PAH}$, has been calculated using the code of Visser et al. (in prep), following the 
method of \cite{Draine01}. This opacity includes a graphite-like continuum opacity and 
is similar to that of \cite{Li01} (see their
Fig. 2). For this paper, we opted to use a single generic PAH molecule consisting
of neutral C$_{50}$H$_{20}$. In reality, the PAH component is much more complicated and
requires the use of a full chemical code to determine. For the purposes of fitting the broad-band
photometry of the nebulosity around VV Ser, a simple PAH opacity is deemed sufficient. 
The model framework developed for VV Ser is a good starting point for including
a full PAH chemical network. 
 
\begin{figure}
  \plotone{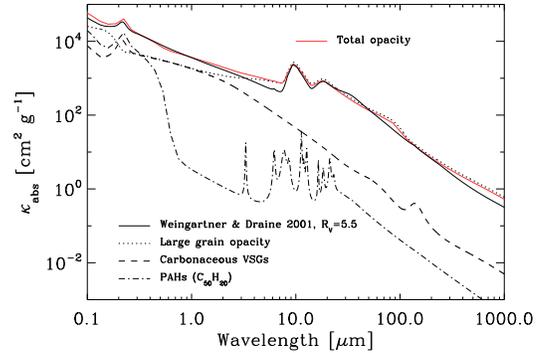}
  \caption{The opacities used for the envelope. The PAH and VSG opacities has been scaled to 5\% and 6\% of the total dust mass, respectively. 
The solid red curve is the sum of the opacities. The total opacity is compared to that of
\cite{Weingartner01} for $R_V=5.5$, which also takes PAH and VSGs into account. }
  \label{opacities}
\end{figure}

The nature of the VSG component is even less constrained. It is clear that a quantum-heated component
is needed to produce the 24--70\,$\mu$m flux. To achieve this, the grains must be heated by
UV--near-infrared photons to temperatures of order 150--300\,K. If UV photons are responsible, then 
grains significantly larger than the PAHs are needed. In the literature, both silicate and carbonaceous
VSGs have been proposed, presumably as amorphous solids to avoid producing strong spectral resonances. 
Following \cite{Weingartner01}, we assume that the VSGs are mostly carbonaceous in nature, with a size distribution corresponding
to a small grain tail of the larger grains. For the VSGs, we use the opacities of spherical graphite grains in the Rayleigh limit. 
The enthalpy curve for the VSGs is the same as that of the PAHs with
a single size of 500 C atoms. The exact choice in size
makes little difference in the absorption coefficient, but is important when calculating the heat capacity for
a single grain; smaller grains can be heated to higher temperatures by a photon of a given energy. Since the SED of
the nebulosity is not well-constrained in the far-infrared, the size of the VSGs cannot be directly constrained. 
For the PAH/VSG enthalpy curve, we use the parametrized form of
\cite{Chase85}, relevant for graphite. 
The properties of the quantum heated grains are summarized in Table \ref{parameters} and the opacity curves of the various envelope dust
components are compared in Fig. \ref{opacities}.

\section{Fitting the envelope structure}
\label{Features}

The constructed model is able to fit data spanning four orders of magnitude
in wavelength and five order of magnitude in spatial scale (see Paper I for a discussion of the fit on small
spatial scales). In this section, 
it is discussed how each observational property is connected to the model.

\begin{table}
\centering
\caption{VV Ser envelope model parameters}
\begin{tabular}{lll}
\hline
\hline
Parameter&Model value\\
\hline
$\rho_{\rm out}(5\times 10^4\,\rm AU)$ &$500\pm 200$\,cm$^{-3}$&\\
$R_{\rm in}$ & 15\,000\,AU\\
$R_{\rm out}$ & 50\,000\,AU\\
PAH size&C$_{50}$H$_{20}$&\\
VSG size&500\,C\\
$A_{\rm PAH}$&$5\pm 2$\%$^1$ \\
$A_{\rm VSG}$&$6\pm 3$ \%$^{1,2}$ \\
Disk shadow pos. angle& $15\pm5\degr$\\
Disk incl. & $72\pm5\degr$\\
\hline
\end{tabular}
\begin{itemize}
\item[$^1$] Abundances are given in terms of percentage of the total dust mass (quantum heated + thermalized grains).
\item[$^2$] The VSG abundance is for the south-eastern clump.
\end{itemize}
\label{parameters}
\end{table}

\begin{figure}
  \plotone{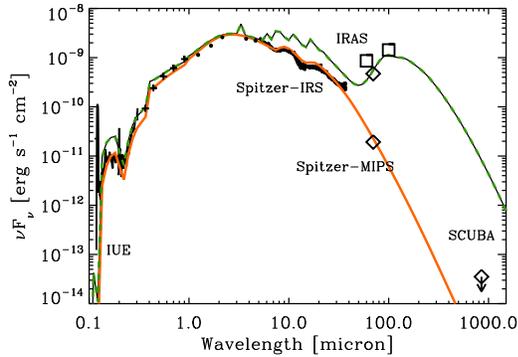}
  \caption{Spectral energy distribution of VV Ser in a 50,000\,AU aperture (dashed green curve) and a 600\,AU aperture (full red curve) 
the quantum-heated nebulosity. The $\Box$ symbols indicate IRAS fluxes from the IRAS extended source catalog, while the
$\Diamond$ symbols indicate Spitzer-MIPS fluxes for the entire nebulosity (upper) and the central VV Ser point source (lower). 
The JCMT-SCUBA point indicates an upper limit for a point source at the location of VV Ser. 
The model SED including the quantum-heated nebulosity has been calculated by
integrating the flux over the south-eastern lobe within 50,000\,AU, where most of the envelope emission is, and subtracting
the background cloud emission measured in an annulus spanning 50,000--100,000\,AU.}
  \label{sed_vsg}
\end{figure}

\subsection{Mid-infrared nebulosity}
\label{nebulosity}
A central property of the model of VV Ser is the ability to reproduce the 
bright nebulosity observed at mid-infrared wavelengths with Spitzer-IRAC and MIPS. Clearly, 
a mid-infrared nebulosity extending over $6\arcmin$ associated with a young star is not a common
occurrence. Assuming a distance of 260\ pc mid-infrared emission is detected up to 47,000\,AU from the central
star. In IRAC4, MIPS1 and MIPS2 (centered on 24\,$\mu$m and 70\,$\mu$m) 
a maximum surface brightness of 3, 4 and 18 mJy/sq. arcsec is reached at a distance of 15,000\,AU
(see Fig. \ref{cross}). This corresponds to color temperatures between IRAC4 and MIPS1 of 360\,K and between MIPS1 and MIPS2 of 85\,K.
Tests using the model showed that it is not possible for dust grains in thermal equilibrium with the radiation
field from the central star to produce such bright nebulosity in the mid-infrared at these radii, except at 70\,$\mu$m where
small thermal grains can explain the emission. The inclusion
of quantum-heated grains readily produces a nebulosity with high surface brightness around a star 
as luminous as VV Ser. A possible way of producing non-thermal continuum emission is 
via quantum-heated VSGs or PAHs or a mix of both. In this context it is important to note that the nebulosity is bright at
all wavelengths beyond $\sim$5.6\,$\mu$m (IRAC 3) and in fact continues to rise to dominate the total SED at 70\,$\mu$m.
It is also interesting to note that the 24 and 70\,$\mu$m emission is not smoothly distributed around the source, but is strongly
concentrated $50\arcsec$ to the south-east of the star. The 8\,$\mu$m emission, on the other hand, appears more
evenly distributed. All imaging modes in which the nebulosity is detected show a dark band centered on the star with a 
position angle of 10--20\degr.  

\begin{figure}
  \plotone{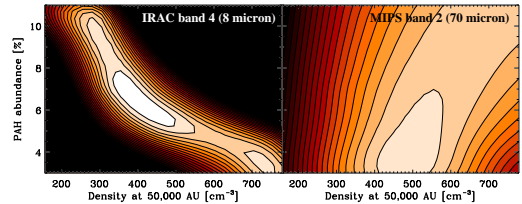}
  \caption{Goodness of fit (see text) as a function of PAH abundance in
percent of the the total dust mass and cloud density. The contours
have been normalized, since the substructure in the nebulosity renders an absolute 
$\chi^2$ value meaningless. The contour levels are plotted in steps of 15\%. 
The lighter contours indicate a value of 1 and the darkest 3. 
The constraints used are the IRAC 4 and MIPS 1 bands
together with the IUE spectrum. The fit
is slightly degenerate in the sense that a higher density may be coupled
with a lower PAH abundance, but only as long as the cloud does not approach unity optical
depth for UV photons within the nebula. This happens for densities of $\gtrsim 2000\,\rm cm^{-3}$. }
  \label{CHI2}
\end{figure}

\begin{figure*}
  \plottwo{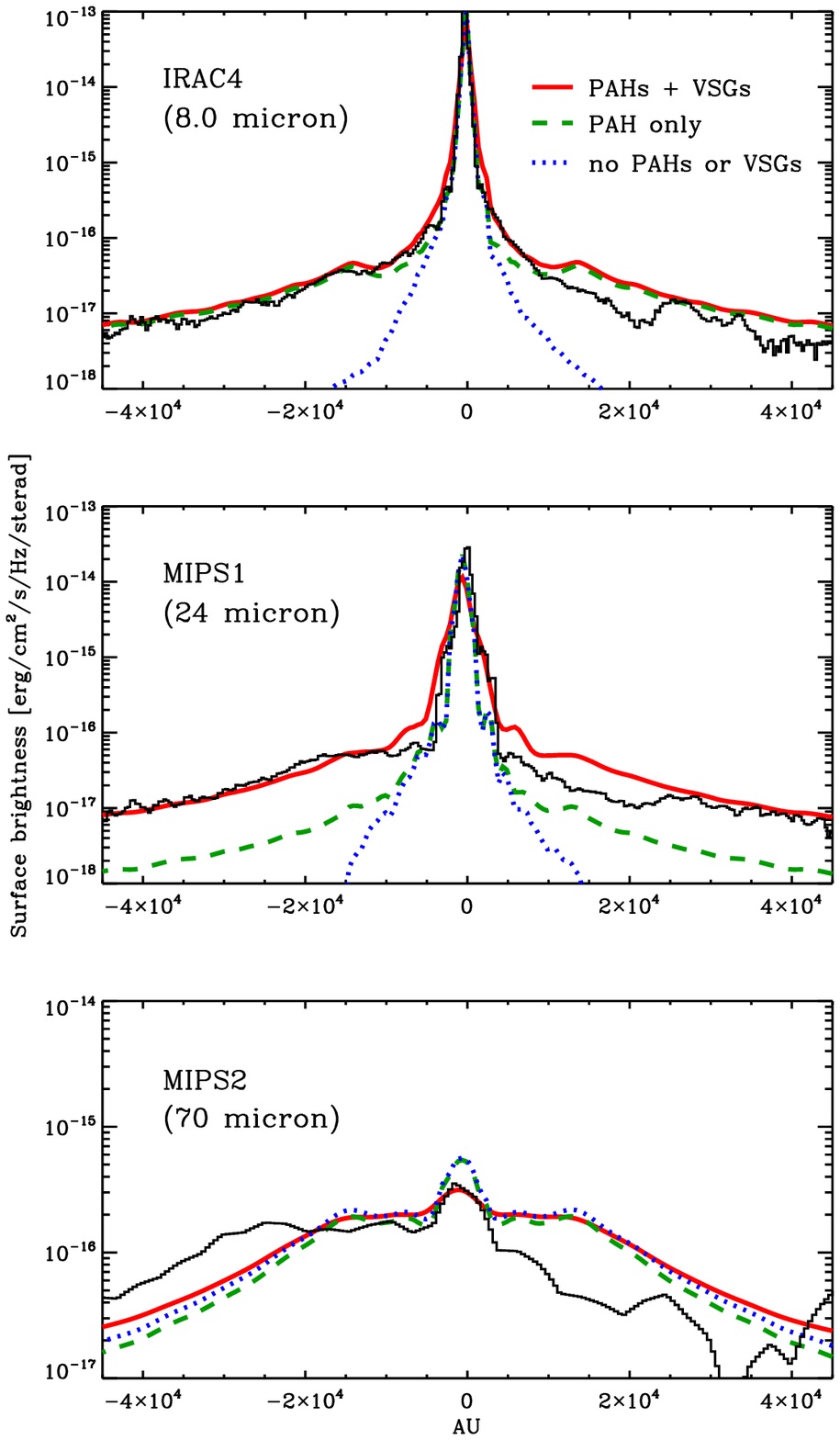}{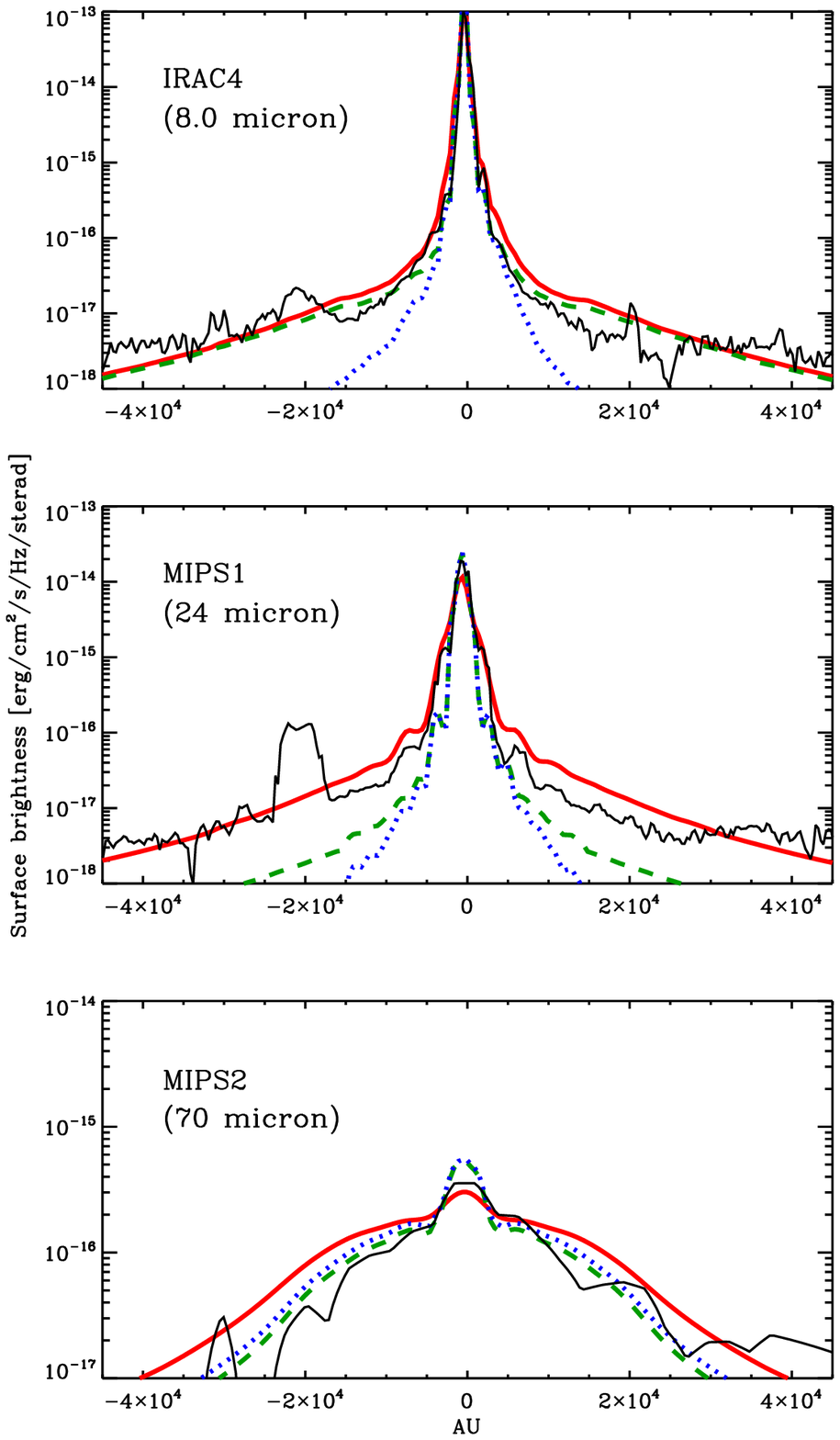}
  \caption{Cross sections through VV Ser in the Spitzer pass bands oriented perpendicular (left column; positive direction to the west) 
and parallel (right column; positive direction to the south) to the disk plane. The solid black curves are the observations, while the dashed, green curves are for the
model with the VSGs removed, and the dotted blue curves are for models with both PAHs and VSGs removed. The solid, red curves are those of the best-fit model, including 
both PAHs and VSGs.
The model including VSGs is only 
substantially different from the PAH only model outside the disk shadow and at wavelengths longer than 
$\sim$15\,$\mu$m. At 70\,$\mu$m, thermal grains dominate the emission within 20,000\,AU of the central star. 
 }
  \label{cross}
\end{figure*}

The observed nebulosity clearly has a complex 3-dimensional structure that 
cannot be modeled in detail by an axi-symmetric code. The model is intended to
fit an average surface brightness. However, we argue that the mid-infrared emission must be optically thin, and that 
the surface brightness therefore is a function of column density and UV field only. This allows the model to be used
to fit an average PAH abundance in the surrounding cloud. The quantum-heated PAHs are included in the model setup as 
described in Sect. \ref{QH}. The surrounding material is assumed to have a density profile
decreasing as $R^{-1}$, except within 15,000\,AU of VV Ser, where the cloud density is constant and lowered by a factor of 10 relative to the density at 15,000\,AU to fit the fact that
the nebulosity peaks at 15,000\,AU. In addition to the spherical cavity, the morphology of the nebulosity suggests the presence of a conical cavity perpendicular to
the disk, reminiscent of an outflow cavity. The envelope is thus modeled by the following expression for $R>$15,000\,AU:
\begin{equation}
\rho = C(\theta,R) \times  (R/R_{\rm out})^{-1},
\end{equation} 
where:
\begin{equation}
C(\theta, R) = \left\{
  \begin{array}{cc}
    \rho_{\rm out} & \mbox{for } \theta \geq (1-R/R_{\rm out}) *\pi/3.5\\
    \rho_{\rm out}/4 & \mbox{for } \theta < (1-R/R_{\rm out}) *\pi/3.5
  \end{array}
\right.
\end{equation}

The parameters of the cavity were not varied in the fit.
The envelope emission can then be modeled using four parameters: the total
dust density at 15,000\,AU, the PAH and VSG abundances and the total extinction toward the central star. The extinction or total column density of the cloud
is considered fixed. This leaves three parameters to be varied
to fit the surface brightness profiles at 8.0, 24 and 70\,$\mu$m, where the nebulosity is detected with sufficient signal-to-noise. 

To find the best-fitting cloud density and PAH abundance, we calculated a grid of models, comparing each model
to the imaging data. The model images were convolved with theoretical Spitzer point spread functions, calculated using the STINYTIM package ver. 1.3 for 
pre-launch parameters (Krist 2002, SSC web site). The goodness-of-fit estimate is
the sum of the squared differences in pixel values between the observed and model intensities in a cross section along the polar axis: $\Sigma[I_{\rm obs}(y)-I_{\rm model}(y)]^2$.

We find that by using only the
IRAC 4 band for the goodness-of-fit estimate, the model becomes degenerate in the sense that a lower cloud density can be compensated
by a higher PAH abundance. This is simply the optically thin limit. However, since the relevant optical
depth is measured in the UV region of the spectrum, increasing the density too much will cause the 
nebulosity to become optically thick to UV photons at relatively small distances to the central star.
In the model grid, we find that emission above 50\,$\mu$m, and therefore also the MIPS2 70\,$\mu$m emission, is dominated by the thermal grains, rather than emission from the
VSGs. This provides a convenient independent measure of the density of the thermal grains.
Therefore, by comparing the 8, 24 and 70\,$\mu$m bands, the PAH and VSG abundances can be constrained 
to $5\pm 2$\% and $6\pm 3$\% of the total dust mass. The outer cloud density is found to be 
$\rho_{\rm out}=500\pm200\,\rm cm^{-3}$. The result is illustrated by the goodness-of-fit surfaces in Fig. \ref{CHI2}. The best-fitting 
PAH abundance corresponds
to $2.5\pm0.8\times 10^{-5}\,\rm H^{-1}$ or $12\pm 4$\% of the C atoms being bound in PAHs assuming a gas-to-dust 
ratio of 100 and a cosmic C/H ratio of $2\times 10^{-4}$ \citep{Holweger01,Allende02}.
This is consistent with other estimates of the abundance of PAHs in molecular clouds and 
photo-dissociation regions \citep{Boulanger98,Tielens99}.
Note that the cloud density matches well that ($\sim 700\,\rm cm^{-3}$) 
determined using the extinction map of \cite{Cambresy99}.

Fig. \ref{cross} shows model cross sections of the best-fitting model compared to the Spitzer images 
parallel and perpendicular to the disk plane of VV Ser. 
Model cross sections for envelopes with no VSGs and with no quantum heated grains are also shown. 
This plot illustrates the constraints used to fit the envelope. The addition of the PAH component is clearly
necessary to fit the 8\,$\mu$m image, while the VSGs are required to fit the 24\,$\mu$m image. The 70\,$\mu$m
image shows significant extended emission due to thermal grains. The model somewhat under-predicts the surface brightness
in the western part of the envelope at this wavelength, although that could possibly be corrected by changing the VSG opacity. 

\begin{figure*}
\centering
\includegraphics[width=5.2cm]{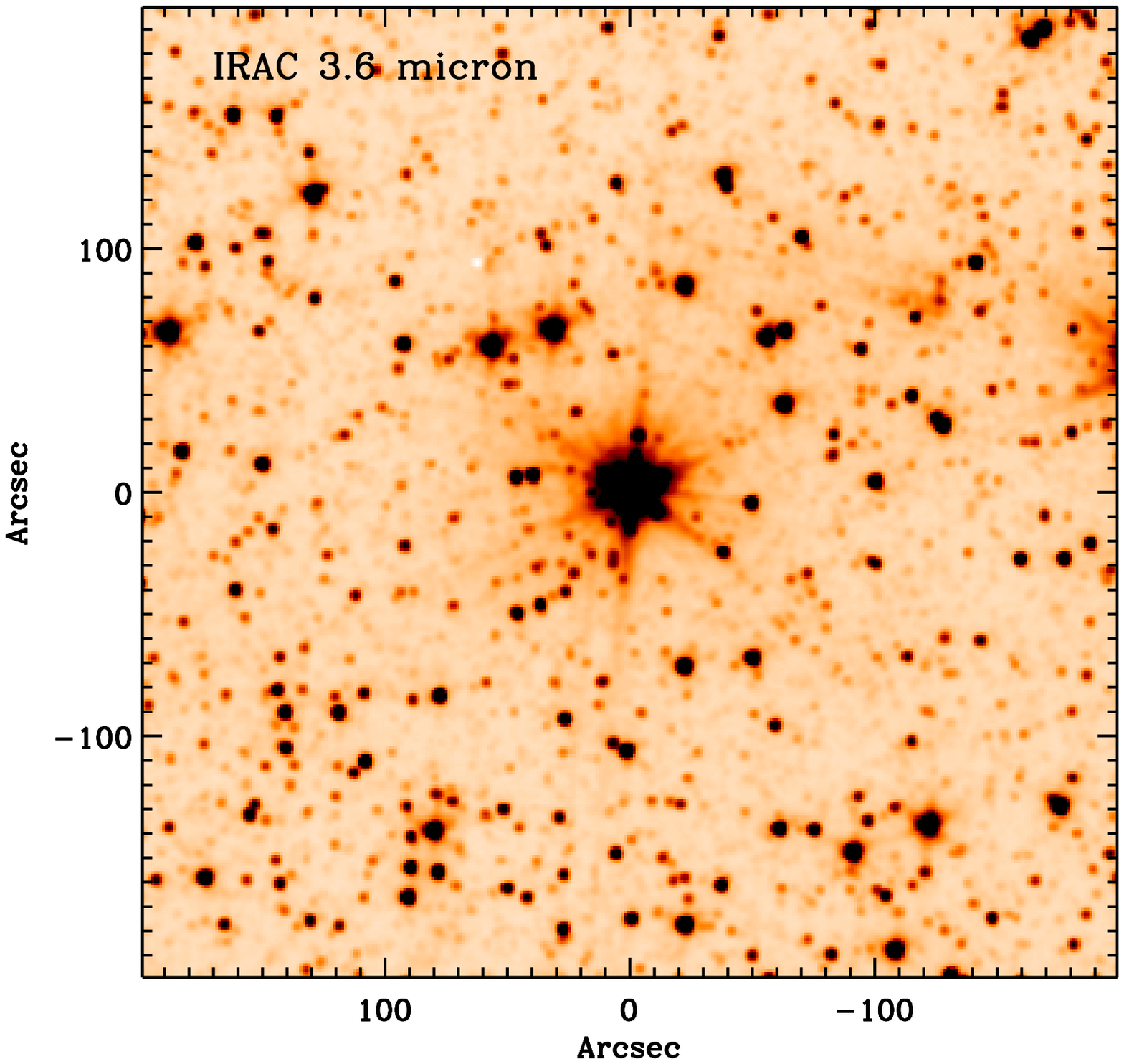}
\includegraphics[width=5.2cm]{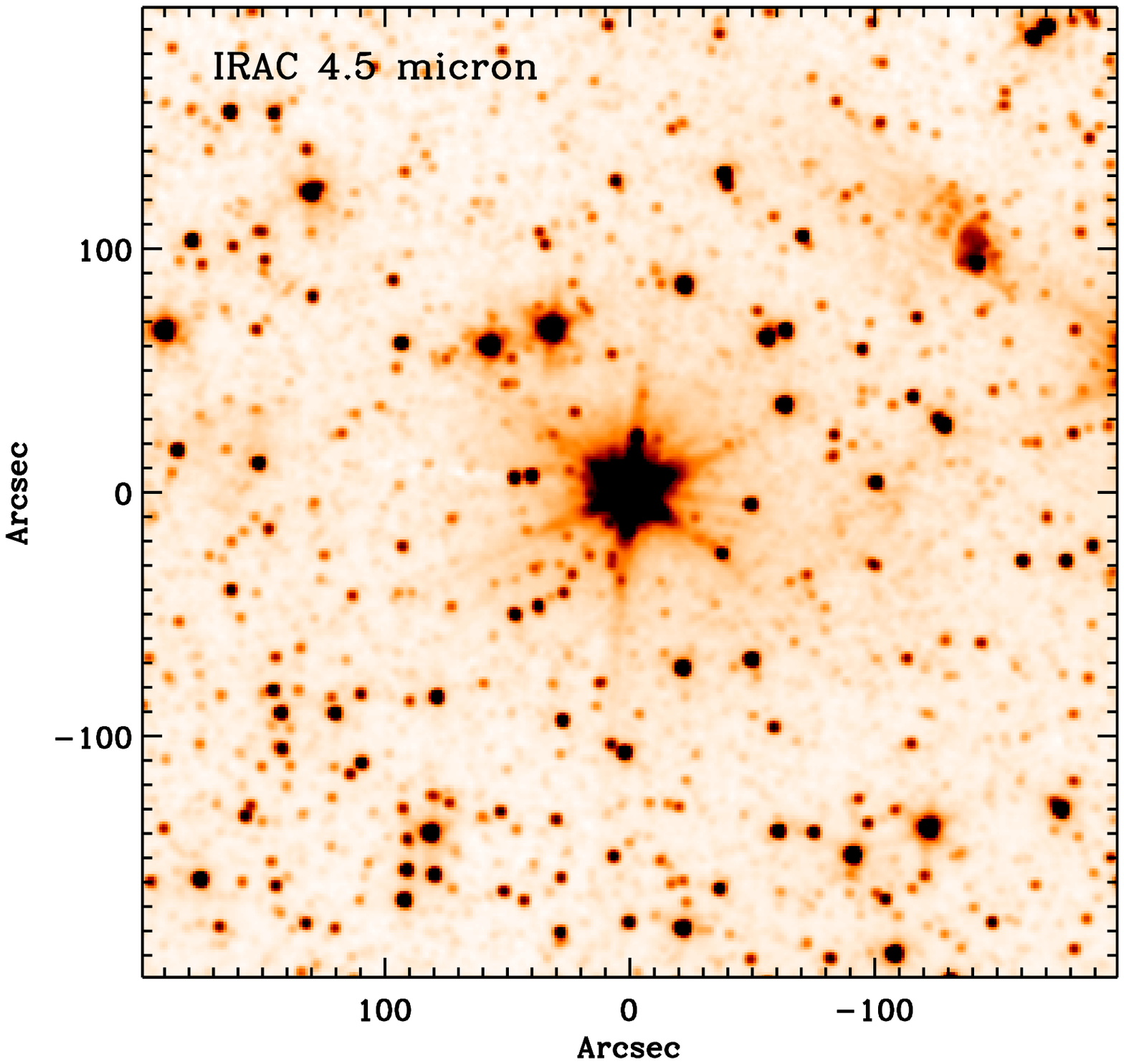}
\includegraphics[width=5.2cm]{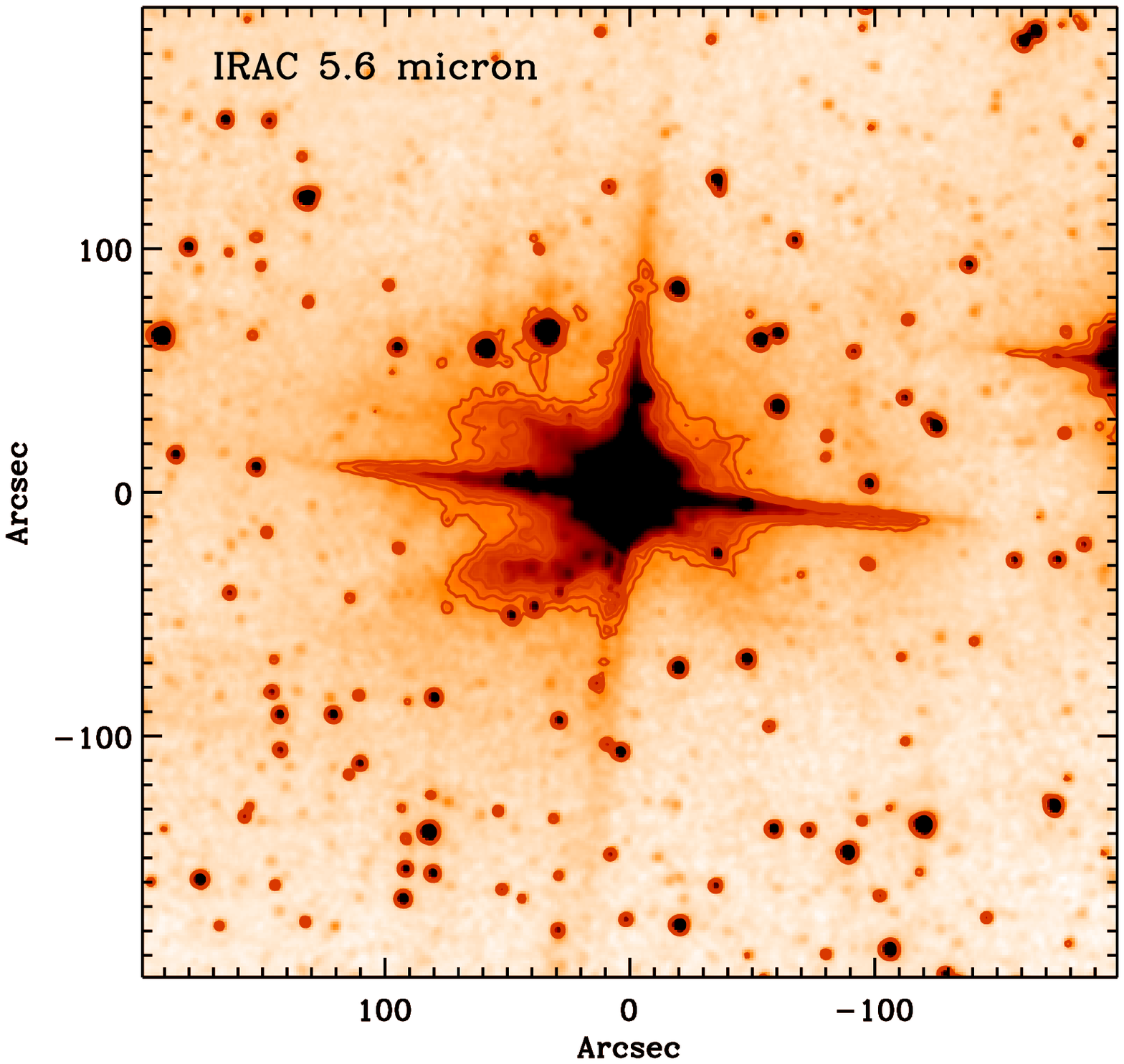}
\includegraphics[width=5.2cm]{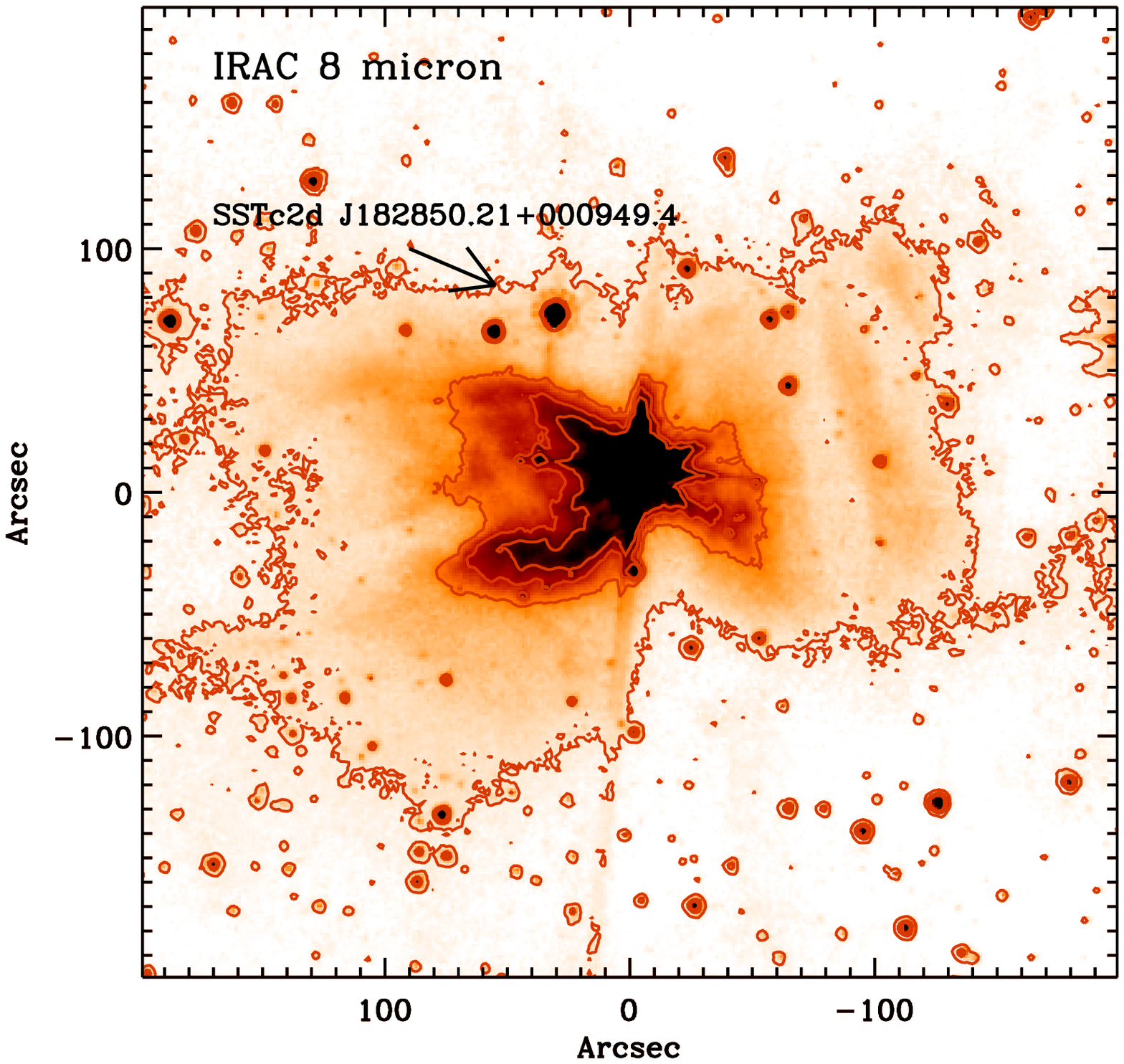}
\includegraphics[width=5.2cm]{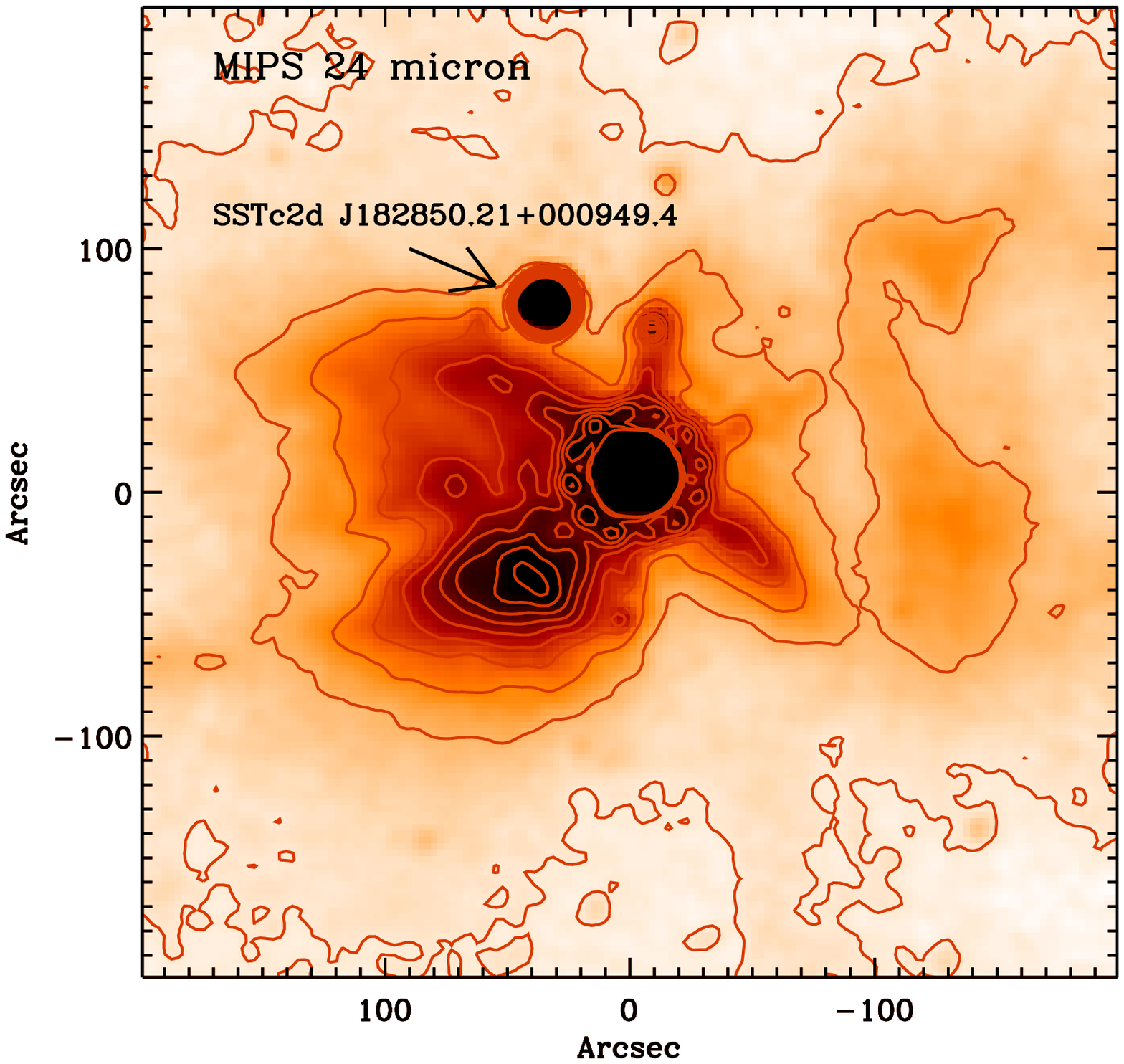}
\includegraphics[width=5.2cm]{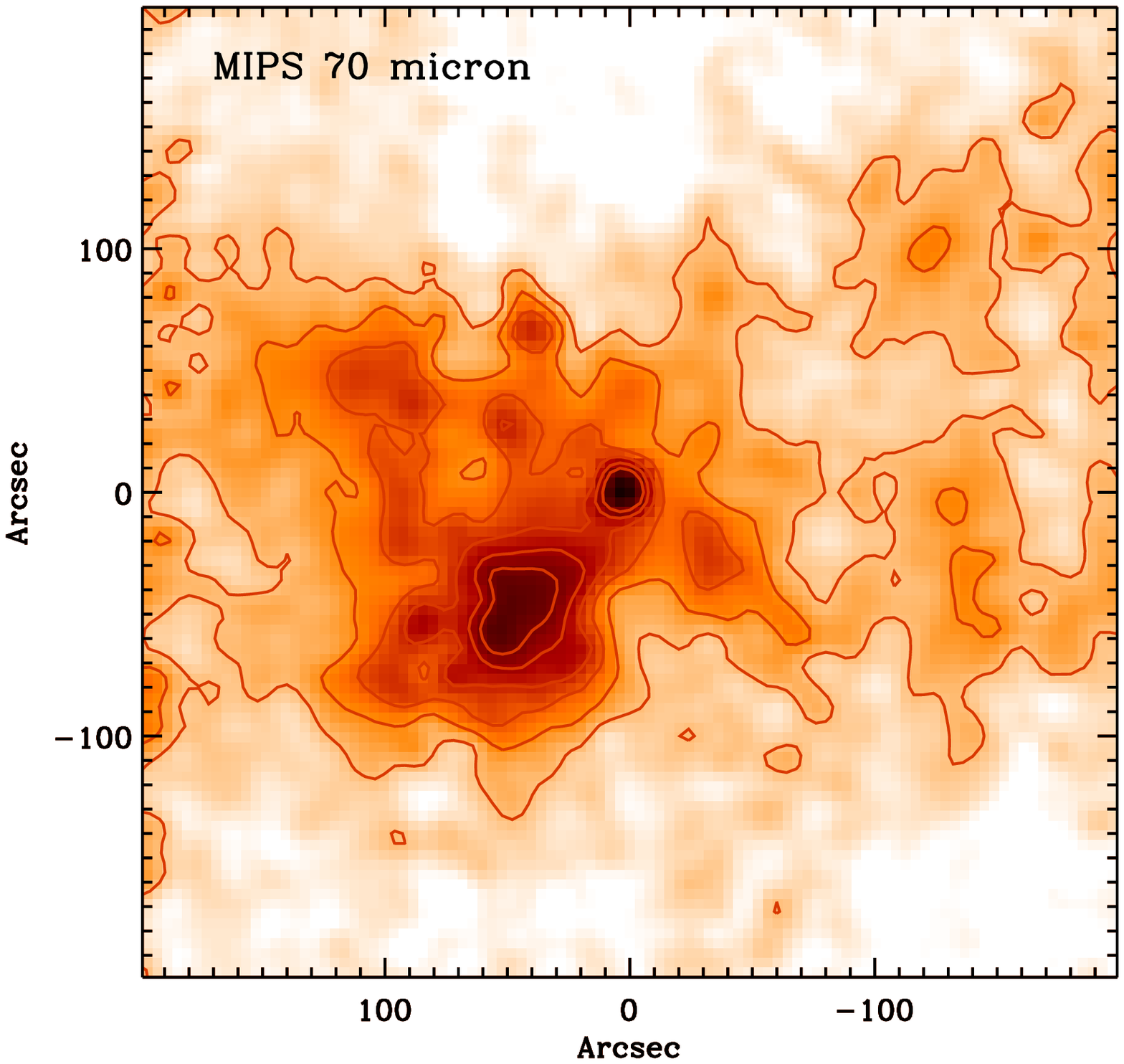}
\includegraphics[width=5.2cm]{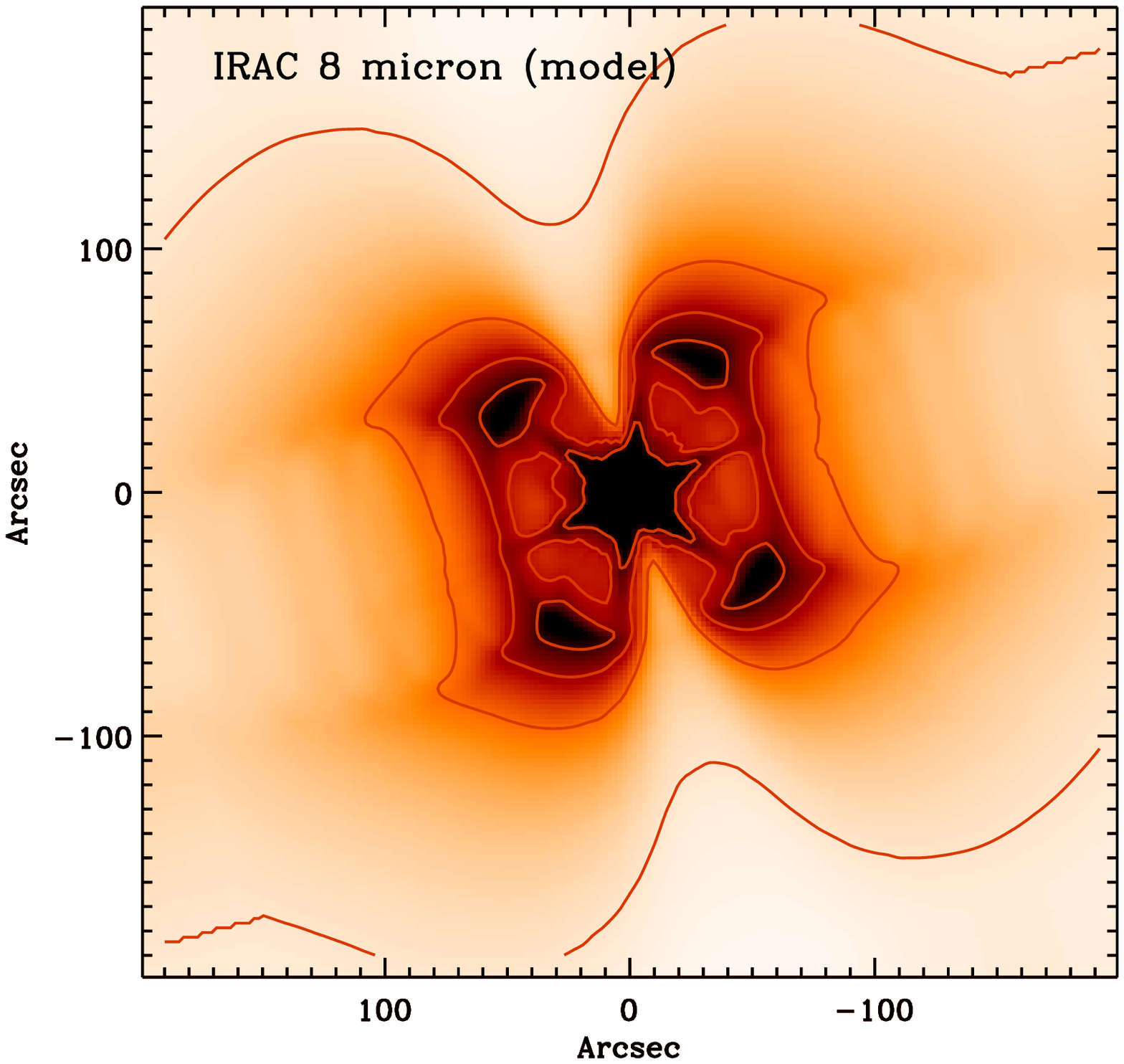}
\includegraphics[width=5.2cm]{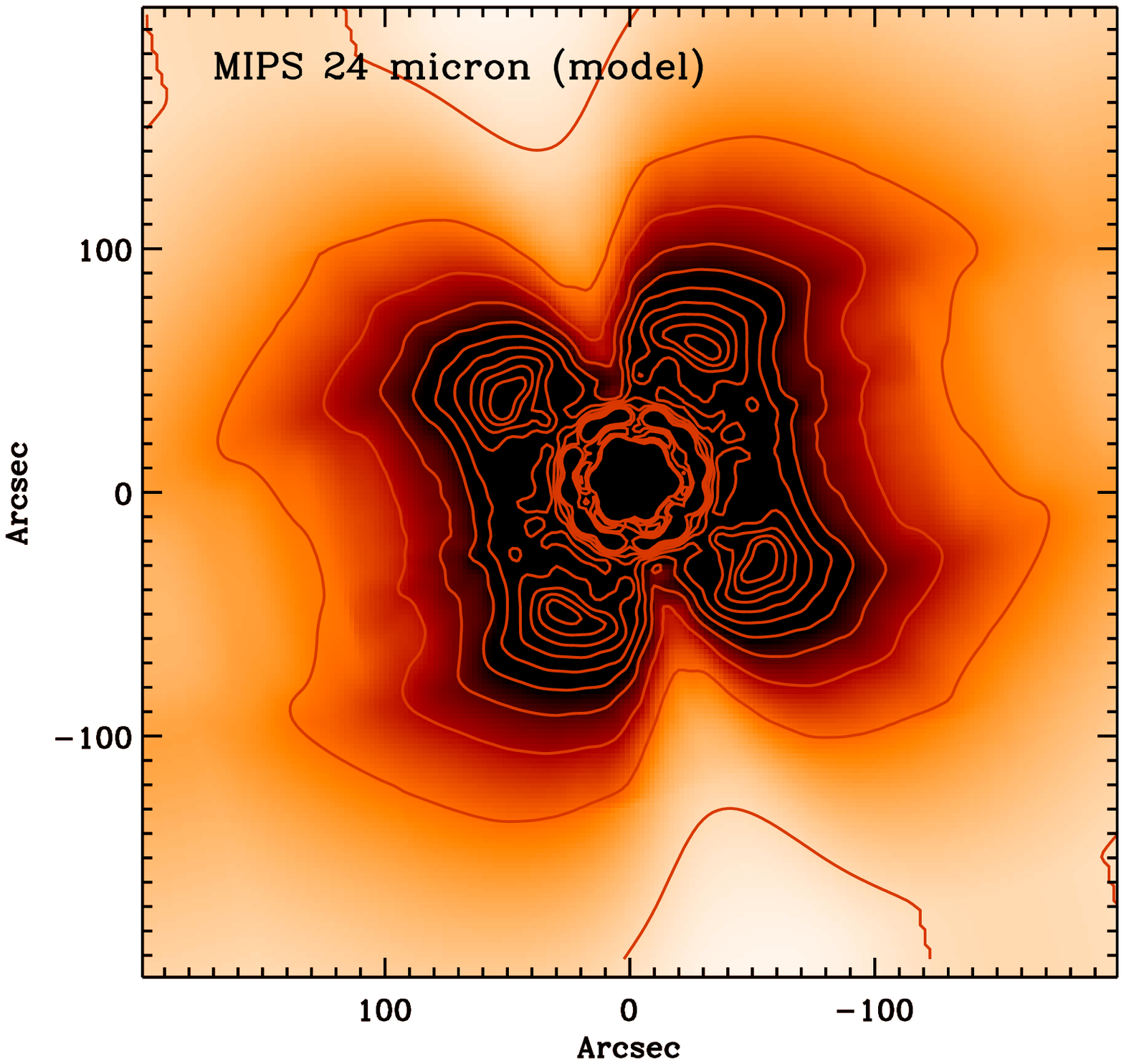}
\includegraphics[width=5.2cm]{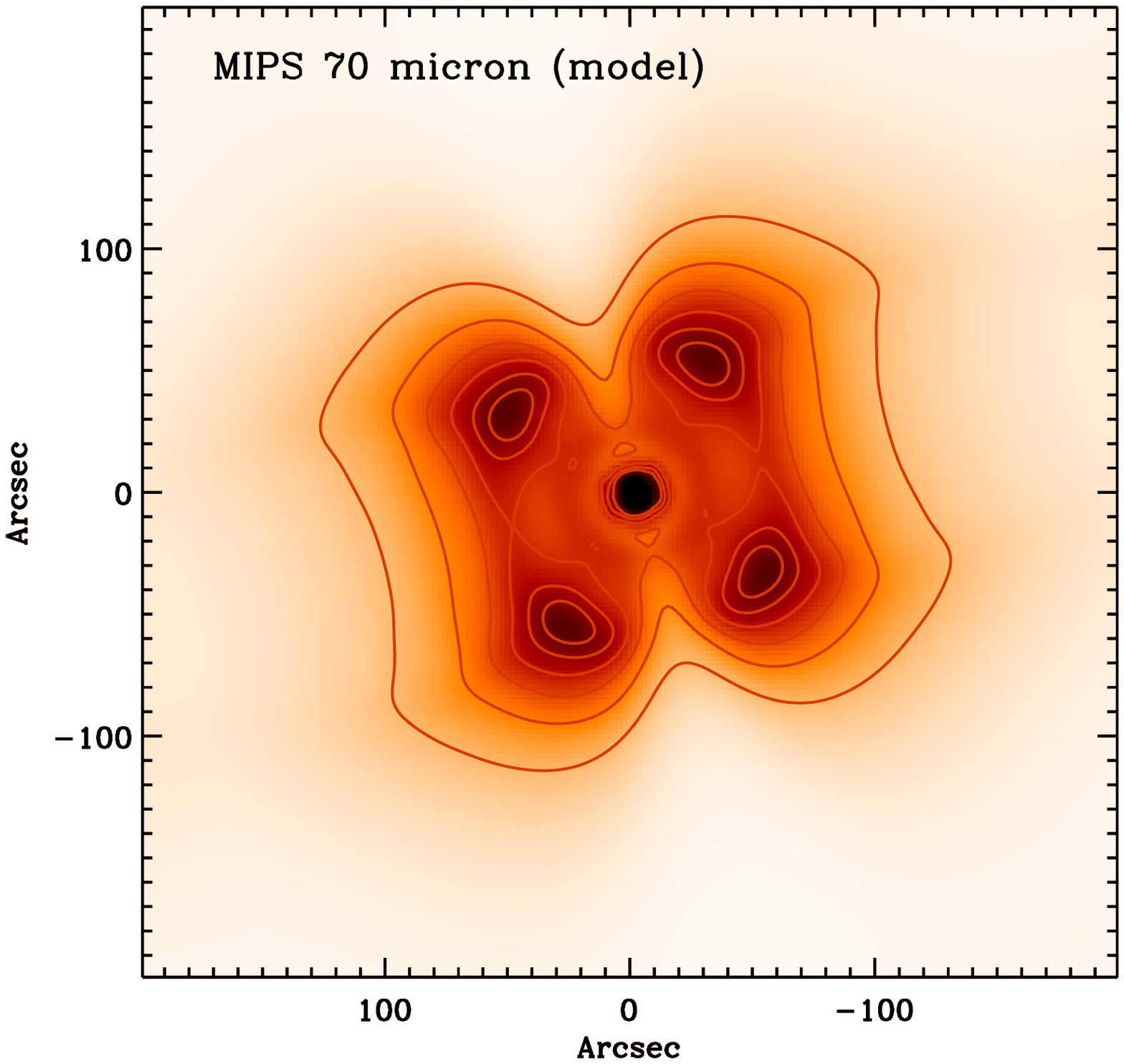}

  \caption{{\it Left panels:} IRAC4 (8\,$\mu$m), MIPS1 (24\,$\mu$m) and MIPS2 (70\,$\mu$m) images of the VV Ser
region. The IRAC4 image is dominated by PAH emission and background stars, while the MIPS images
are dominated by emission due to a population of VSGs. The nearby red object SSTc2d J182850.21+000949.4 is not
related to VV Ser and is possibly a chance projection of a T Tauri star outside the main cloud or a background
AGB star. {\it right panels:} Model contours in the IRAC4, MIPS1 and MIPS2 photometric bands. The contours correspond to those
of the left side panels. The IRAC4 model contours are overlaid on the SOFI-NTT $J$-band image to show that 
there are no visible signatures of the mid-infrared nebulosity in the near-infrared. 
North is up and East is to the left. }
  \label{images}
\end{figure*}

Fig. \ref{images} shows the Spitzer images from 4.5 to 70\,$\mu$m, along with the model fits at 8, 24 and 70\,$\mu$m. 
The IRAC and MIPS images are seen to be reasonably well fitted in terms of the shadow and the south-western emission blob. 
However, the axisymmetric nature of the model makes the fit worse regarding other details in the morphology of the envelope emission.

\subsection{Disk shadow}

A wedge-like dark band in the mid-infrared nebulosity appears naturally in the model and matches well that 
observed in the IRAC and MIPS images.  
This is due to the disk blocking UV photons from propagating through the
disk plane causing a shadow to be cast across the PAH nebulosity where the small grains do not get 
quantum-heated because of the absence of UV photons. 
While the nebulosity and superposed shadow are expected to be similar in morphology to those seen in reflection nebulae, 
this is a new 
type of shadow governed by an entirely different physical process from single scattering. The opening angle of the shadow is
a measure of the maximum height $H_{\rm UV}/R$ of the $\tau_{\rm UV}=1$ surface of the disk. 
In the case of VV Ser, the shadow is wide, corresponding to a $H_{\rm UV}/R_{\rm UV}\sim 0.3$. For 
a self-shadowed disk like VV Ser, the maximum $H_{\rm UV}/R_{\rm UV}$ is located at the puffed-up inner rim. 
The height at which the inner rim becomes optically thick to UV photons is therefore $\sim0.25$\,AU at the inner rim
$R_{\rm in}=0.85$\,AU. The inner rim scale height achieving this is about 50\% larger than that expected from a
disk in hydrostatic equilibrium. A scale height smaller than that of an equilibrium disk
would be easier to explain, for instance as being due to dust settling. A
larger scale height is more difficult to understand. 
Note, however, that the height at which the disk becomes optically thick to UV photons
is degenerate in the scale height and the total disk mass, with a larger disk mass compensating for a smaller scale height. 
A significant change in the disk mass of 1-2 orders of magnitudes is required to achieve this effect.
Alternatively, since the shadow is only probing the vertical density profile at large heights, a relatively small change in
the wings of the profile at large heights while preserving a hydrostatic (Gaussian) profile in the disk mid-plane may explain the observations. 
One possibility for such a departure is that small grains are lifted to large heights by the flow of a disk wind \citep{Shu94}. Detailed
calculations will be required to determine whether this process can produce a significant dust column density along the line of sight. 

Also evident is that the shadow is not straight in the {\it Spitzer} images, i.e. does not follow a straight line through
the origin, but curves away from this line. For a disk shadow, this effect occurs when the disk inclination deviates
significantly from edge-on and allows the inclination of VV Ser to be constrained to $70-75\degr$ 
\citep[see also the discussion in][]{Pontoppidan_shadow}. The model also reproduces the ratio of surface brightness inside the
shadow, as compared to that out of the shadow, which is an interesting test of the density structure of the disk model. 

While the shadow appears in the images because 
quantum-heated grains are excited by high energy photons only, the lack of UV photons in the disk plane is expected to 
significantly affect the chemistry of the molecular cloud as well. In other words, a different chemistry is
expected to be seen inside the shadow as compared to outside [see also \cite{Semenov05}].   

\subsection{Near-infrared images}
The near-infrared images from the NTT-SOFI are useful for obtaining an upper limit to the surface brightness 
of any scattered light at these wavelengths. It is important that this is consistent with the cloud density
derived from the mid-infrared nebulosity. The observed upper 
limit on the surface brightness is $\sim$22 and 20.5\ mag/sq. arcsec for the $J$ and $Ks$ bands, respectively. 
The model yields a surface brightness due to scattering of $J=23.5$, $H=23.1$, and 
$K=22.9$\ mag/sq. arcsec at $20\arcsec$ from the star if the quantum-heating is turned off, i.e. these values refer
to pure scattering. However, including the
quantum-heated PAHs increases both the $J$- and $K$-band surface brightnesses by 2 magnitudes. 
This is due to the near-infrared continuum opacity introduced to the PAH opacity in \cite{Li01} to fit observations of
reflection nebulae in massive star-forming regions. 
This means that slightly deeper imaging is expected to show nebulosity in the near-infrared.
In this context, it is also interesting to note that the increase of 2-3 orders of magnitude 
in the near-infrared PAH opacity compared to the \cite{Li01} opacity as suggested by \cite{Mattioda05} 
would result in easily detectable extended emission in the SOFI images. This this is not seen, suggesting
that the PAH opacity curve in the VV Ser nebulosity is similar that of \cite{Li01} and that the PAHs are
therefore likely mostly neutral or negatively charged. Alternatively, significantly larger PAHs than those 
used here may also reduce the near-infrared extended emission. 

\subsection{UV spectrum}
A high quality ultra-violet spectrum of VV Ser obtained with the International Ultraviolet Explorer (IUE)
exists (see also Paper I). This provides an important direct constraint on the UV power injected into the system. The 
UV radiation field in the envelope and the resulting PAH excitation 
can therefore be calculated with some confidence. Care has been taken in fitting the SED to the UV spectrum. 
It is interesting to note that the IUE spectrum shows shallow
absorption from the well-known, but unidentified, $2175\,\rm \AA$ band always seen in the galactic interstellar 
extinction 
curve (see \cite{Draine03} and references therein). 
Our model of VV Ser produces a 2175\,\AA feature 2 times deeper than that observed, indicating that this band is too strong in the assumed PAH
opacity. 

Large PAH molecules have previously been identified as a possible
carrier of the $2175\,\rm \AA$ band \citep{Weingartner01} (see also Fig. \ref{opacities}). However, these studies
fixed the PAH abundance to fit the observed $2175\,\rm \AA$ band. Since the emission nebula surrounding VV Ser 
provides an independent
measure of the PAH abundance that produces a $2175\,\rm \AA$ band that is even too strong, our results seem to support that the PAHs
likely contribute significantly to this feature.
No good fit can be obtained to the IUE spectrum
unless the $2175\,\rm \AA$ feature is at least dominated by the PAH opacity. For example, if the $2175\,\rm \AA$ feature
is dominated by another carrier, the PAH abundance would be lowered significantly (by at least a factor of 4). 
To compensate, a fit to the 8\,$\mu$m
nebulosity would require the density to be increased beyond the $\sim 1000\,\rm cm^{-3}$ where the $\tau_{\rm UV}\sim 1$ 
surface passes inside the observed quantum heated nebulosity. It should be noted that since the actual UV opacity of PAHs 
is not well known, significant uncertainties remain. In particular, if the actual PAH opacity does not have a $2175\,\rm \AA$ feature,
contrary to the \cite{Li01} opacity, then no strong constraints on the PAH abundance can be derived from the UV spectrum. 

\section{Discussion}
\label{Discussion}

How common are extended quantum-heated nebulosities associated with single (low- or intermediate-mass) 
young stars and what do they tell us? 
In this study, the mid-infrared nebulosity around VV Ser is interpreted as belonging to the general 
Serpens molecular cloud, i.e. the observed PAH/VSG emission is consistent with a constant density cloud. The carrier
of the far-infrared emission, on the other hand, is not distributed evenly. 
As such the nebulosity is a direct measure of the UV radiation field around 
VV Ser; the nebulosity exists because the density (and therefore PAH density) is not so low as to not have sufficient 
PAH molecules, yet the density is not so high as to block all UV photons from penetrating to the observed 
edge of the nebulosity at 30,000\,AU. At the same time, there is a strict upper limit on the density 
of small dust grains surrounding VV Ser 
from the lack of reflection nebulosity in the near-infrared. The lack of extended emission in the near-infrared
also puts an upper limit of the near-infrared opacity of the PAHs, roughly consistent with the opacity
of \cite{Li01}. Deeper optical/near-infrared imaging may detect the faint 
reflection nebulosity predicted to be there. In the case of VV Ser, this translates to a gas density 
of $\sim 500\,$cm$^{-3}$, assuming a dust-to-gas ratio of 100 - a value that corresponds well to that
derived by the extinction map of \cite{Cambresy99}. If the density is increased by a factor
of a few, the PAH nebulosity around VV Ser becomes significantly more compact, though still visible, because of extinction of UV photons
by the large grain population. 
It is therefore expected that Spitzer will detect more of such quantum-heated nebulae associated with stars that happen
to be embedded in the more diffuse parts of molecular clouds.
 
Such nebulae may be important for several reasons. First, a well-defined stellar photosphere gives the 
possibility for accurately calculating the input radiation field; the input spectrum can be determined by simply
obtaining a UV-optical spectrum of the star in question. In the absence of a star one must resort to
using a much more poorly defined interstellar radiation field. Using a radiative transfer model coupled with
a spectrum of the nebulosity, it is possible to directly measure the emissivity of the PAH molecules
in the cloud for comparison with laboratory data. For instance, in the case of VV Ser, it can
be concluded that the near-infrared PAH opacity is low, in accordance with the opacity model of \cite{Li01}. 
Furthermore, by mapping differences in PAH features in the
radial direction (through a diminishing radiation field), it is possible to map the chemistry, hydrogenation 
and ionization state of the PAHs and couple those directly to the radiation field. Finally, the far-infrared spectral
slope of the nebulosity will constrain the composition and size distribution of the VSG component. Such sources may 
therefore be very important for deciphering the role of PAHs and VSGs in molecular clouds. 

The fact that in our self-shadowed disk model  
most of the observed disk shadow is created by the innermost parts of the disk suggests that
the mid-infrared nebulosity may be variable on timescales of 1-10 years (at least the dynamical timescale of the 
puffed-up inner rim). This possible variability should be seen as variations in the shadow opening angle. A 
long term variability of VV Ser has indeed been observed in the optical, in addition to the UX Orionis events (see Paper I), which 
may suggest that this is reflected also in the shadow morphology. However, it is highly unlikely that the shadow
will experience strong changes, such as disappearing entirely.  

\section{Conclusions}

{\it Spitzer} imaging of the Herbig Ae/UX Orionis star VV Ser has shown it to be surrounded by a very
extended, but bright nebulosity. The nebulosity is detected from IRAC band 3 at 5.6\,$\mu$m to MIPS2 at 70\,$\mu$m. 
Using the disk model for VV Ser developed in Paper I as a starting point, we model the extended mid-infrared
nebulosity surrounding the star by adding small transiently heated grains to the surrounding material.

\begin{itemize}
\item We show that the nebulosity is most likely due to quantum-heated small grains consisting of both PAHs and somewhat larger
carbonaceous/silicate grains of perhaps $\sim 500$ atoms. The best-fitting abundances are 4-6\% and 5-7\% by mass for PAHs and VSGs, 
respectively, although these values depend somewhat on the opacities used. Assuming the PAH opacity of \cite{Li01} is appropriate, 
we find that the $2175\,\rm \AA$ feature must be mostly due to the graphite-like structures within the PAH molecules. 

\item While the PAHs appear to be evenly distributed at 
distances of 15,000--50,000\,AU from the central star, the VSG component is more structured with most of this material concentrated
in a clump $\sim50\arcsec=12,500\,$AU to the south-east of VV Ser. 

\item The nebulosity is observed to be bisected by a wedge-shaped
dark band extending across the entire nebulosity with a position angle of $\sim 10-20\degr$ (east of north). We interpret
this as a large shadow cast by the small $\sim$50\,AU central disk by blocking UV photons from reaching the surrounding cloud. This
is fully consistent with the disk model developed in Paper I. The opening angle of the shadow shows that the opening angle
of the $\tau_{\rm UV}=1$ surface of the disk is $H/R\sim 0.3$. The inferred disk inclination of $72\pm5 \degr$ and position angle
of ($13\pm 5\degr$ east of north)
are important constraints for interpreting the model presented in Paper I, in particular in relation to the
near-infrared interferometry of \cite{Eisner04}. 

\item We propose that the VV Ser nebulosity and other isolated quantum heated regions excited by a single star with
a well-defined UV spectrum may be used to strongly constrain the properties of PAHs and other types of small grains
in molecular clouds. For instance, the relation of the ionization state of PAHs to the UV field can be accurately
mapped with a mid-infrared spectral map of the nebula. 

\item Additionally, the shape of the mid to far-infrared spectrum of the 
nebula will constrain the size distribution of VSGs. In principle, such observations coupled with simple radiative transfer
calculations can serve as a calibration for studies of more complex regions containing quantum heated grains, such
as extra-galactic sources, massive star-forming regions and circumstellar disks. 

\item Disk shadows cast on
quantum heated nebulosities are useful for measuring the $\tau_{\rm UV}=1$ surface of circumstellar disks and therefore
provide constraints on the grain size distribution in the upper layers of disks. We have also shown that due to the
large size of disk shadows, they are very convenient for determining inclination and position angles for disks that
are otherwise impossible to spatially resolve through direct imaging. Spitzer-IRAC and MIPS imaging surveys are
excellent tools for identifying quantum heated nebulosities around single stars, and Spitzer-IRS is well suited for
follow-up spectroscopy. 
\end{itemize}

\acknowledgements{The authors wish to thank Ruud Visser for use of his code to calculate the PAH opacity and Jean-Charles Augereau
for useful discussions. The referee, Kenneth Wood, is thanked for comments that significantly improved the quality of the manuscript.
Support for this work was provided by NASA through Hubble Fellowship grant \#01201.01 awarded by the Space Telescope Science Institute, which is operated by the Association of 
Universities for Research in Astronomy, Inc., for NASA, under contract NAS 5-26555.
Astrochemistry in Leiden is supported by a SPINOZA grant of the Netherlands Organization for Scientific Research (NWO).
Support for this work, part of the Spitzer Space Telescope Legacy Science Program, was provided
by NASA through Contract Numbers 1224608 and 1230779 issued by the Jet Propulsion Laboratory, California Institute of
Technology under NASA contract 1407. This research was supported by the European Research Training Network ``The Origin of Planetary System'' (PLANETS, contract HPRN-CT-2002-00308). }

\bibliographystyle{apj}
\bibliography{ms}

\end{document}